\documentclass[showpacs,amsmath,amssymb,twocolumn,nofootinbib]{revtex4}
\begin{document}
\title{Tolman-Oppenheimer-Volkoff equations in presence of
the Chaplygin gas: stars and wormhole-like solutions}

\author{V. Gorini}
\affiliation{Dipartimento di Scienze Fisiche e
Mathematiche, Universit\`a dell'Insubria, Via Valleggio 11, 22100
Como, Italy
\\INFN, sez. di Milano, Via Celoria 16, 20133 Milano, Italy}
\author{
A. Yu. Kamenshchik}
\affiliation{Dipartimento di Fisica and INFN, via Irnerio 46, 40126 Bologna, Italy
\\L.D. Landau Institute for Theoretical Physics,
Russian Academy of Sciences, Kosygin str. 2, 119334 Moscow, Russia}
\author{U. Moschella}
\affiliation{Dipartimento di Scienze Fisiche e
Mathematiche, Universit\`a dell'Insubria, Via Valleggio 11, 22100
Como, Italy
\\INFN, sez. di Milano, Via Celoria 16, 20133 Milano, Italy}
\author{V. Pasquier}
\affiliation{Service de Physique Th\'eorique, C.E. Saclay, 91191
Gif-sur-Yvette, France}
\author{A. A. Starobinsky}
\affiliation{L.D. Landau Institute for Theoretical Physics,
Russian Academy of Sciences, Kosygin str. 2, 119334 Moscow, Russia}

\begin{abstract}
We study static solutions of the Tolman--Oppenheimer--Volkoff
equations for spherically symmetric objects (stars) living in a
space filled with the Chaplygin gas. Two cases are considered. In
the normal case all solutions (excluding the de Sitter one) realize
a three-dimensional spheroidal geometry because the radial
coordinate achieves a maximal value (the "equator"). After crossing
the equator, three scenarios are possible: a closed spheroid having
a Schwarzschild-type singularity with  infinite blue-shift at the
"south pole", a regular spheroid, and a truncated spheroid having a
scalar curvature singularity at a finite value of the radial
coordinate. The second  case arises when the modulus of the pressure
exceeds the energy density (the phantom Chaplygin gas). There is no
more equator and all solutions have the geometry of a truncated
spheroid with the same type of singularity. We consider also static
spherically symmetric configurations existing in a universe filled
with the phantom Chaplygin gas only. In this case two classes of
solutions exist: truncated spheroids and solutions of the wormhole
type with a throat. However, the latter are not asymptotically flat
and possess curvature singularities at finite values of the radial
coordinate. Thus, they may not be used as models of observable
compact astrophysical objects.
\end{abstract}

 \maketitle

\section{Introduction}

Because of the nowadays accepted existence of cosmic
acceleration \cite{cosmic,dark}, the study of spherically
symmetric solutions of the Einstein equations \cite{Tolman,OV}
in presence of dark energy is of much interest. This study has
already been undertaken for instance in
\cite{cosmstar,chapstar,darkstar,BS2007}.

One of the simplest models for dark energy is the Chaplygin gas
 \cite{we,chap}. The model is based on a
perfect fluid satisfying the equation of state $p = -A/\rho$,
where $p$ is the pressure, $\rho$ is the energy density and $A$
is a positive constant. Some studies have already appeared
where the problem of finding spherically symmetric or
wormhole-like solutions of Einstein's equations in presence of
the Chaplygin gas have been addressed \cite{chapstar,Kuh}.

Here we study static solutions of the Tolman--Oppenheimer--Volkoff
(TOV) equations for spherically symmetric objects living in a space
filled with the Chaplygin gas. Results obtained appear to be very
different from the apparently similar problem of stars in presence of a
cosmological constant \cite{cosmstar}. Indeed, in the latter case
the exterior solution of the TOV equations is nothing but the
well-known Schwarzschild-de Sitter geometry, while the interior
problem essentially coincides with the standard TOV case. The only
difference is that the pressure does not vanish at the star
surface; on the contrary, it is negative and its absolute value is
equal to the cosmological constant.

Instead, the Chaplygin gas strongly feels the presence of the star,
and consequently the solution acquires quite unusual features. These
features are the existence of a maximal value of the radial
coordinate, dubbed ``equator'', and the appearance of
curvature singularities at some finite values of the radial
coordinate $r$. We find also that for the case of the phantom Chaplygin
gas when the absolute value of the pressure is greater than the
energy density ($|p| > \rho$) wormhole-like solutions with a throat
exist. However, these solutions cannot be identified with the
usual Morris-Thorne-Yurtsever wormholes \cite{worm-stand} because
they are not asymptotically flat. Moreover, they possess curvature
singularities at finite values of $r$ as well.

The structure of the paper is as follows: in
Sec. II we write down the TOV equations in presence of the Chaplygin
gas and describe two exact solutions of them; Sec. III is devoted to
the analysis of the normal case where $|p| < \rho$ while in Sec. IV
we study  the phantom case with $|p| > \rho$, and  we consider the
star-like configurations analogous to those studied in Sec. III; in
Sec. V we consider the solutions of the TOV equations existing in a
universe filled exclusively with the phantom Chaplygin gas. Sec. 6
contains conclusions and discussion.

\section{Tolman-Oppenheimer-Volkoff equations in the presence of
the Chaplygin gas}

We suppose that the universe is filled with a perfect fluid
with energy-momentum $ T_{\mu\nu} = (\rho + p)u_{\mu}u_{\nu} -
g_{\mu\nu}p\label{emt} $ and consider a static spherically
symmetric interval
\begin{equation}
ds^2 = e^{\nu(r)} dt^2 - e^{\mu(r)}dr^2 -r^2(d\theta^2 +
\sin^2\theta d\phi^2). \label{metric}
\end{equation}
Then the Einstein system reduces to the
following pair of equations:
\begin{equation}
e^{-\mu}\left(\frac{1}{r}\frac{d\mu}{dr} - \frac{1}{r^2}\right)
+\frac{1}{r^2} = 8\pi\rho, \label{tt} \end{equation}
\begin{equation}
e^{-\mu}\left(\frac{1}{r}\frac{d\nu}{dr} + \frac{1}{r^2}\right)
- \frac{1}{r^2} = 8\pi p, \label{rr} \end{equation}
plus the
energy-momentum conservation equation
\begin{equation}
 \frac{dp}{dr} = -\frac{d\nu}{dr} \frac{\rho+p}{2}.
 \label{conserv1}
\end{equation}
Solving Eq. (\ref{tt}) with the boundary condition $e^{-\mu(0)}
= 1$ gives
\begin{equation}
e^{-\mu} = \left(1 - \frac{2M}{r}\right) \label{mu}
\end{equation}
where, as usually,  $M(r)= 4\pi \int_0^r dr r^2 \rho(r).
\label{mass} $ This is equivalent to
\begin{equation}
\frac{dM}{dr} = 4\pi r^2 \rho,\ M(0) = 0. \label{mass1}
\end{equation}
Eq. (\ref{rr}), (\ref{conserv1}) and (\ref{mu}) together give
rise to the TOV differential equation \cite{Tolman,OV}
\begin{equation}
\frac{dp}{dr} = -\frac{(\rho+p)(M+4\pi r^3
p)}{r(r-2M)}.\label{TOV}
\end{equation}
Complementing  equations (\ref{mass1}) and (\ref{TOV}) with an
equation of state relating  $p$ and $\rho$ one has a closed system
of three equations for the three variables $p,\rho$ and $M$. In
this paper we investigate the case when the fluid is  the
Chaplygin gas whose equation of state is
\begin{equation}
p = -\frac{\Lambda^2}{\rho}. \label{Chap}
\end{equation}
Then Eqs. (\ref{mass1})  and  (\ref{TOV})  give rise to the
following system of first-order differential equations for $p$ and
$M$:
\begin{eqnarray}
\frac{dp}{dr} &=& \frac{(\Lambda^2-p^2)(M+4\pi r^3
p)}{pr(r-2M)},\label{TOV1}\\
\frac{dM}{dr} &=& -\frac{4\pi \Lambda^2 r^2}{p}. \label{mass11}
\end{eqnarray}
We denote the radius of the star by $r_b$. As usual we suppose that
the pressure is continuous at the surface of the star. The
``exterior'' problem amounts to considering system (\ref{mass11}) in
the interval $r> r_{b}$ with some properly chosen boundary
conditions $p(r_{b})$ and $M(r_{b})$ at $r=r_{b}$. It is easy to see
that at $r > r_b$ the system admits two exact solutions with
constant pressure. The first of them
\begin{equation}
p = -\rho = -\Lambda,
\ \ \ \
M = \frac{4}{3}\pi \Lambda r^3, \label{dS1}
\end{equation}
describes the geometry of the de Sitter space with
\begin{equation}
e^{\nu} = e^{-\mu} = 1 - \frac{r^2}{r_{dS}^2}, \label{dS2}
\;\;\;
r_{dS} = \sqrt{\frac{3}{8\pi\Lambda}}.
\end{equation}
The second solution is the Einstein static universe
\begin{equation}
p = -\frac{\Lambda}{\sqrt{3}},~~\rho=\sqrt{3}\Lambda,~~
M = \frac{4\sqrt{3}\pi
\Lambda r^3}{3} \;\;\; \label{ES}
\end{equation}
with the radius
\begin{equation}
r_{E} = \sqrt{\frac{\sqrt{3}}{8\pi\Lambda}}=\frac{r_{dS}}
{3^{1/4}}.\label{EC2}
\end{equation}

\section{The normal case: $|p| < \rho$}

We now consider solutions with a non-constant pressure. Some
additional constraints have to be imposed on the boundary
conditions:
\begin{equation}
-\Lambda < p(r_{b}) < 0,\label{constr}
\end{equation}
\begin{equation}
M(r_{b}) < \frac{r_{b}}{2}.\label{constr1}
\end{equation}
First of all, note that the pressure $p$ cannot attain the values
$p = 0$ and $p=-\Lambda$ in the region where $2M(r) < r$.
Indeed, for $ 2M(r) < r$ the right-hand side of (\ref{TOV1}) is
negative, while in order to approach $p = 0$ starting from
negative values of $p$, it is necessary to have $dp/dr> 0$. In
addition, let us rewrite Eq. (\ref{TOV1}) as follows:
\begin{equation}
d\ln(\Lambda^2 - p^2) = -2 dr \frac{(M+4\pi r^3
p)}{r(r-2M)}.\label{pL}
\end{equation}
and suppose that $p \rightarrow -\Lambda$  as   $r \rightarrow
r_1$,   with  $r_1
> 2M(r_1)$. Then, upon integration of Eq. (\ref{pL}), we get a
divergence at the left-hand side and a regular expression at the
right-hand side, a  contradiction.

Thus, as long as the condition $2M(r) < r$ is satisfied  we have
$-\Lambda < p(r) < 0$ and $\rho(r) > \Lambda$. This means that the
mass $M(r)$ is growing at least as fast as  $r^3$,  so that at
some radius $r = r_0$ the  equality
\begin{equation}
M(r_0) = \frac{r_0}{2} \label{hor1}
\end{equation}
is achieved. Then at $r=r_0$ we must have  $p(r_0) = p_0 =
-\frac{1}{8\pi r_0^2}$. Indeed, let us expand the relevant
quantities around $r_0$:
\begin{equation}
r = r_0 -\varepsilon, \label{vicin}
\end{equation}
\begin{equation}
M(r) = \frac{r_0}{2} - \tilde{M}(\varepsilon) \label{vicin1}
\end{equation}
\begin{equation}
p(r) = p_0 + \tilde{p}(\varepsilon), \label{vicin2}
\end{equation}
where $\tilde{M}(\varepsilon)$ and $\tilde{p}(\varepsilon)$
tend to zero when $\varepsilon \rightarrow 0$. Equation
(\ref{TOV1}) has the following asymptotic form:
\begin{equation}
\frac{d\tilde{p}}{d\varepsilon} = \frac{(\Lambda^2-p_0^2)(1 + 8\pi
r_0^2 p_0)}{2\varepsilon (p_0 + 8\pi r_0^2
\Lambda^2)},\label{TOVas}
\end{equation}
from which it is easily seen that $ \tilde{p} \sim \ln \varepsilon$
when $\varepsilon\rightarrow 0$, unless  $p_0 = -\frac{1}{8\pi r_0^2}$.

We now determine a lower and an upper bound  for the radius $r_0$.
First, note that since $p_0 > - \Lambda$ we have
\begin{equation}
r_0 > \sqrt{\frac{1}{8\pi\Lambda}}.
\label{limit}
\end{equation}
On the other hand, since $\rho(r) > \Lambda$,  Eq. (\ref{mass1})
implies  that of $M(r) > \frac{4\pi \Lambda r^3}{3}$ so that
\begin{equation}
 r_0 < \sqrt{\frac{3}{8\pi\Lambda}}.
\label{limit1}
\end{equation}

The asymptotic equation for $\tilde{p}$ has the form
\begin{equation}
\frac{d\tilde{p}}{d\varepsilon} = \frac{\tilde{p}}{2\varepsilon} +
C_0 \label{TOVas1}
\end{equation}
where
\begin{equation}
C_0 = \frac{1}{8\pi r_0^3}\left(\frac32 - 32\pi^2\Lambda^2
r_0^4\right),\label{C0}
\end{equation}
and its  solution is
\begin{equation}
\tilde{p} = A \sqrt{\varepsilon} + 2C_0 \varepsilon
\label{sol-as}
\end{equation}
where $A$ is an arbitrary coefficient. Thus, the family of the
solutions $p(r), M(r)$ can be characterized by the two parameters
$r_0$ and $A$ which, in turn, are determined by the boundary
conditions $M(r_b), p(r_b)$ on the surface of the star.

However, the coordinates which we have used so far are not
convenient for the problem under consideration since the metric
coefficient $g_{rr} = e^{\mu} = (1-2M/r)^{-1}$ has a fictitious
(coordinate) singularity at $r=r_0$. Therefore, instead of the
coordinate $r$, we introduce a new
coordinate $\chi$ defined by  $r=r_0\sin\chi$, so that the
corresponding metric becomes
\begin{equation}
ds^2 = e^{\bar{\nu}(\chi)} dt^2 - e^{\bar{\mu}(\chi)}d\chi^2
-r_0^2\sin^2\chi(d\theta^2 + \sin^2\theta d\phi^2).
\label{metric1}
\end{equation}

Then the  $tt$-component of the Einstein equations  has the form
\begin{equation}
e^{-\bar{\mu}}(\bar{\mu}'\cot \chi + 2 - \cot^2\chi) +
\frac{1}{r_0^2\sin^2\chi} = 8\pi\rho \label{tt1}
\end{equation}
where prime denotes differentiation with respect to the variable
$\chi$. Integration  with the boundary condition
$e^{-\bar{\mu}(0)} = 1/r_0^2$ gives
\begin{equation}
e^{-\bar{\mu}} = \frac{1}{r_0^2\cos^2\chi}\left(1 - \frac{8\pi
r_0^2}{\sin\chi}\int_0^{\chi}\rho(\chi)\sin^2\chi\cos\chi
d\chi\right).\label{mu1}
\end{equation}

 From Eqs. (\ref{hor1}) and (\ref{mass}) it follows that
\begin{equation}
r_0^2 = \frac{1}{8\pi \int_0^{\pi/2} \rho(\chi)\sin^2\chi\cos\chi
d\chi} \label{r0choice}
\end{equation}
that implies the positivity and finiteness of the expression
(\ref{mu1}).

For latter purposes we also write down  the $\chi\chi$ component
of the Einstein equations:
\begin{equation}
e^{-\bar{\mu}}(\cot^2\chi + \bar{\nu}'\cot\chi) -
\frac{1}{r_0^2\sin^2\chi} = 8\pi p.\label{rr1}
\end{equation}

The energy-momentum  conservation equation now reads
\begin{equation}
\bar{\nu}' = -\frac{2p'}{p+\rho}, \label{conserv2}
\end{equation}
and for the case of the Chaplygin gas  it can be easily integrated
to give
\begin{equation}
e^{\bar{\nu}} = \frac{A_0}{\Lambda^2 - p^2} \label{nu}
\end{equation}
where $A_0$ is some positive constant which fixes the choice of
the time scale.

Combining Eqs. (\ref{tt1}), (\ref{rr1}) and (\ref{conserv2}) one
gets
\begin{equation}
p' = -\frac{\cos\chi(p+\rho)(\bar{M} + 4\pi r_0^3 \sin^3\chi p)}
{\sin\chi(r_0\sin\chi-2\bar{M})} \label{TOVchi}
\end{equation}
where
\begin{equation}
\bar{M}(\chi) = 4\pi r_0^3\int_0^{\chi} \rho \sin^2\chi \cos\chi
d\chi \label{Mbar}
\end{equation}
and
\begin{equation}
\bar{M}' = 4\pi r_0^3 \rho\sin^2\chi\cos\chi,~\bar{M}(0) = 0~.
\label{Mbar1}
\end{equation}
Relation (\ref{hor1}) can be rewritten as
\begin{equation}
r_0 = 2\bar{M}(\pi/2), \label{c0choice1}
\end{equation}
and for the Chaplygin gas Eqs. (\ref{TOVchi}) and (\ref{Mbar1})
acquire the forms
\begin{equation}
p' = \frac{\cos\chi(\Lambda^2-p^2)(\bar{M} + 4\pi r_0^3 \sin^3\chi
p)} {p\sin\chi(r_0\sin\chi-2\bar{M})},\label{TOVchi1}
\end{equation}
\begin{equation}
\bar{M}' = -\frac{4\pi r_0^3 \Lambda^2\sin^2\chi\cos\chi}{p}.
\label{Mbar2}
\end{equation}
respectively.

We study these equations  in the vicinity of the ``equator''
($\chi = \pi/2$) by introducing a small positive variable $\alpha$
such that
\begin{equation}
\chi = \frac{\pi}{2} - \alpha, \label{alpha}
\end{equation}
and the functions $\tilde{p}(\alpha)$ and $\tilde{M}(\alpha)$:
\begin{equation}
p = -\frac{1}{8\pi r_0^2} + \tilde{p}(\alpha),\label{equator}
\end{equation}
\begin{equation}
\bar{M} = \frac{r_0}{2} - \tilde{M}(\alpha).\label{equator1}
\end{equation}

A simple calculation shows that
\begin{equation}
\tilde{M} = 16\pi^2 r_0^5 \Lambda^2 \alpha^2 + \cdots,
\label{tildeM}
\end{equation}
while for $\tilde{p}(\alpha)$ one can write down the following
asymptotic equation
\begin{equation}
\frac{d\tilde{p}}{d\alpha} = \frac{\tilde{p}}{\alpha} + C_1
\alpha, \label{asympTOV}
\end{equation}
where
\begin{equation}
C_1 = \frac{\frac32 - 32\pi^2 r_0^4\Lambda^2}{8\pi
r_0^2}.\label{C1}
\end{equation}

The solution of Eq. (\ref{asympTOV}) is
\begin{equation}
\tilde{p} = B\alpha + C_1 \alpha^2,\label{asympTOV1}
\end{equation}
with $B$ an arbitrary constant. This solution can be continued to
negative values of the parameter $\alpha$ that corresponds to
the equator crossing. Thus, all  trajectories intersecting
the equator $\chi = \pi/2$ can be characterized by two parameters,
which could be chosen as  $r_0$ and $B$. The static Einstein
solution corresponds to the values $B=0$ and $C_1 = 0$, the latter
condition being equivalent to $r_0^4 = 3/64\pi^2 \Lambda^2$.

In order to investigate  the behavior of the trajectories after
crossing  the equator, we find it convenient to introduce a new
variable
\begin{equation}
y \equiv \frac{1}{\sin\chi},\label{y-define}
\end{equation}
so that $1 \leq y < \infty$. In terms of this variable Eqs.
(\ref{TOVchi1}),(\ref{Mbar2}) can be rewritten as
\begin{equation}
\frac{dp}{dy} = -\frac{(\Lambda^2-p^2)(\bar{M}y^3+4\pi r_0^3
p)}{py^3(r_0-2\bar{M}y)},\label{TOVy}
\end{equation}
\begin{equation}
\frac{d\bar{M}}{dy} =\frac{4\pi \Lambda^2 r_0^3}{py^4}.\label{My}
\end{equation}

Now one can show that the expression $r_0 - 2\bar{M}y$ in the
denominator of the right-hand side of Eq. (\ref{TOVy}) is always
positive at $y > 1$. In order to prove this statement, we first
show that it is true if $p > -\Lambda$.

To this end we introduce the function
\begin{equation}
f(y) \equiv r_0 - r_0 y + \frac{8\pi\Lambda
r_0^3(y^3-1)}{3y^2}~.\label{f}
\end{equation}
Since $p > -\Lambda$, it satisfies the inequality
\begin{equation}
f(y) \leq r_0 - 2\bar{M}(y)y.\label{ineq}
\end{equation}
We have
\begin{equation}
f(1) = 0 \label{initi}
\end{equation}
and
\begin{equation}
f'(y) = -r_0 + \frac{8\pi\Lambda r_0^3}{3} + \frac{16\pi\Lambda
r_0^3}{3y^3}, \label{derf}
\end{equation}
so that
\begin{equation}
f'(1) = r_0(8\pi\Lambda r_0^2 - 1) =
\left(\frac{\Lambda}{|p(r_0)|} -1\right) > 0.\label{derf1}
\end{equation}
Now assume that the function $r_0 - 2\bar{M}(y)y$ becomes equal to
zero at some $y=y_1 > 1$. This means that at some value
$y = y_2 \leq y_1$, the function $f(y)$ vanishes. In turn, this
last condition requires the vanishing of the derivative $f'(y)$ at
some value  $y = y_3 < y_2$. From Eq. (\ref{derf}) one finds
\begin{equation}
y_3 = \left(\frac{16\pi\Lambda r_0^2}{3-8\pi\Lambda
r_0^2}\right)^{1/3}, \label{y3}
\end{equation}
and the requirement $y_3 > 1$ is equivalent to
\begin{equation}
1 < 8\pi\Lambda r_0^2 < 3.\label{ineq3}
\end{equation}

The vanishing of the function $r_0 - 2\bar{M}y$ at the point $y_1$
implies also the vanishing of the expression $\bar{M}y^3 + 4\pi
r_0^3 p$ at this point, i.e. the vanishing of the numerator of the
expression in the right-hand side of Eq. (\ref{TOVy}). Thus, the
values of the functions $\bar{M}$ and $p$ at the point $y_1$ are
given by
\begin{equation}
\bar{M}(y_1) = \frac{r_0}{2y_1},\ \ p(y_1) = -\frac{y_1^2}{8\pi
r_0^2}.\label{y1}
\end{equation}
It follows from the condition $p > -\Lambda$ that
\begin{equation}
y_1^2 < 8\pi\Lambda r_0^2,\label{ineqy1}
\end{equation}
while  from  $y_3 < y_1$ we find
\begin{equation}
y_3^2 < 8\pi\Lambda r_0^2.\label{ineqy3}
\end{equation}
Substituting the expression (\ref{y3}) into the inequality
(\ref{ineqy3}) we get that this inequality is satisfied provided
\begin{equation}
8\pi\Lambda r_0^2 > 4\label{finineq}
\end{equation}
that contradicts the condition (\ref{ineq3}).

Finally, looking at Eq. (\ref{TOVy}) one can see that the pressure
can, in principle, achieve the value $p = -\Lambda$ at some value
$y = y_{\Lambda}$ only if $r_{0}/y_{\Lambda} =
2\bar{M}(y_{\Lambda})$ and $\left(\frac{8\pi\Lambda
r_{0}^2}{y_{\Lambda}^2}-1\right) > 0$. A simple analysis similar
to the one carried out above shows that this is impossible as
well. Thus, we have shown that the expression $r_0 - 2\bar{M}(y)y$
cannot vanish at any value of $y$ in the range $1 < y < \infty$.

We now study the behavior of the pressure at $y > 1$. Here we find
three families of  solutions (geometries). The first one contains
trajectories arriving at the south pole of the
three-dimensional spatial manifold ($y = \infty, \chi = \pi$) with
some value $0 > p(\infty) > -\Lambda$. Looking at Eq. (\ref{TOVy})
we see that the necessary condition for  such solutions to exist
is the convergent behavior of the integral
\begin{equation}
\int dy \frac{\bar{M}y^3+4\pi r_0^3
p}{y^3(r_0-2\bar{M}y)}\label{integral}
\end{equation}
at $y \rightarrow \infty$ that implies  the vanishing of the
function $\bar{M}$ at $y \rightarrow \infty$. Indeed, assume
$\bar{M}(\infty) = M_0 \neq 0$. Then $M_0 > 0$ contradicts the
positivity of the expression $r_0 - 2\bar{M}y$, while $M_0 < 0$
implies the integral (\ref{integral}) to diverge logarithmically.
Thus, the only value $\bar{M}(\infty)$ compatible with $p(\infty)
> -\Lambda$ is $\bar{M}(\infty) = 0$. Then it follows from Eq.
(\ref{My}) that the asymptotic behavior of $\bar{M}$ at $y
\rightarrow \infty$ is
\begin{equation}
\bar{M} = \frac{m}{y^{3}}, \label{asmass}
\end{equation}
where
\begin{equation}
 m = -\frac{4\pi \Lambda^2 r_0^3}{3p(\infty)}~.\label{malpha}
\end{equation}
Substituting the value of $m$  into the integrand of the
right-hand side of Eq. ({\ref{TOVy}), we see that the sign of the
derivative $dp/dy$ at $y \rightarrow \infty$ is determined by the
sign of the expression $(\Lambda^2 - 3p^2(\infty))$. If $p <
-\Lambda/\sqrt{3}$ this derivative is negative, while it is positive
for $p > -\Lambda/\sqrt{3}$.

Thus, there exists a two-parameter family of regular spacetime
geometries for which the spatial manifold represents a
three-dimensional spheroid parameterized by the two parameters
$r_0$ and $p(\infty)$ and the metric coefficient $g_{tt}$  given
by the formula (\ref{nu})  is always positive.

The second family of geometries includes those where the value of
the pressure $p$ becomes equal to zero. Let us describe basic
features of such geometries. We suppose that $p(y_0) = 0$ at $y_0
> 1$. The function $\bar{M}(y)$ cannot
become negative at $y=y_0$ because in this case the derivative
$dp/dy$ would be negative and it would be impossible to reach the
value $p(y_0) = 0$. Hence we consider the case when $\bar{M}(y_0)
= M_0$ where $0< M_0 < \frac{r_0}{2y_0}$. We assume that in the
neighborhood of the point $y_0$ the pressure behaves as
\begin{equation}
p(y) = -D(y_0 - y)^{\alpha} \label{vic0}
\end{equation}
where $D$ and $\alpha$ are some positive constants. Substituting
the expression (\ref{vic0}) into Eq. (\ref{TOVy}), one gets
\begin{equation}
\alpha = \frac{1}{2},\ \ D = \sqrt{\frac{2\Lambda^2 M_0}{r_0 -
2M_0y_0}}. \label{vic01}
\end{equation}
Consider also the case when $\bar{M}(y_0) = 0$. In this case we
look for the expressions describing the behavior of $p$ and
$\bar{M}$ in the vicinity of $y = y_0$ in the form
\begin{equation}
\bar{M}(y) = M_1(y_0-y)^{\beta}, \label{massas}
\end{equation}
\begin{equation}
p(y) = -E(y_0-y)^{\gamma},\label{presas}
\end{equation}
where $M_1$ and $E$ are positive and $0 < \beta < \gamma$.
Substituting expressions (\ref{massas}) and (\ref{presas}) into
Eqs. (\ref{TOVy}), (\ref{My}) one finds the following values for
the parameters $\beta, \gamma, M_1$ and $E$:
\begin{equation}
\beta = \frac13,\ \gamma = \frac23,\ E =
\left(\frac{18\pi\Lambda^4r_0^2}{y_0^4}\right)^{1/3},\ M_1 =
\frac{2r_0}{3\Lambda^2}E^2.\label{param0}
\end{equation}

Note that in this case the values of the of the parameters $M_1$
and $E$ are uniquely fixed by the value of $y_0$. In the case of
$M(y_0) > 0$ considered above one has a one parameter family of
geometries parameterized by the value of $M_0$ or by the value of
$D$. Thus, it seems that one has a three-parameter family of
geometries corresponding to $p \rightarrow 0$ and these parameters
are $r_0, y_0$ and $M_0$. However, Eqs. (\ref{vic01}) or
(\ref{param0}) describe necessary conditions which should be
satisfied to provide the existence of the geometry having the
maximal radius $r_0$ and the pressure $p$ vanishing at $y = y_0$.
Not all the solutions satisfying the relations (\ref{vic01}) or
(\ref{param0}) correspond to ``initial conditions'' at $y =1$,
i.e. $\bar{M}(y=1) = r_0/2$. Moreover, taking into account the
monotonic behavior of the function $M(y)$ one can believe that at
least one value of the parameter $\bar{M}(y_0)$ corresponds to a
geometry with the desirable initial and final conditions. Thus,
the family of solutions (geometries) ending with $p = 0$ is also
two-parametric and can be parameterized by the two parameters
$r_0$ and $y_0$.

An interesting feature of the geometries described above consists
in the presence of the singularity at $y = y_0$. Indeed,
the Chaplygin gas equation of state implies an infinite growth of
the energy density when the pressure tends to zero that in turn
determines the divergence of the scalar curvature. Thus, the
spacetime under consideration cannot be continued beyond
$y = y_0$ or, in other terms, beyond $\chi = \pi - \arcsin
y_0^{-1}$.

The third family of possible geometries includes those for which
the pressure $p$ tends to the value $-\Lambda$ when $y \rightarrow
\infty$ ($\chi \rightarrow \pi$). In this case the acceptable
behavior of the function $\bar{M}$ is $\bar{M}(\infty) = -M_2$,
where $M_2
> 0$. The behavior of the pressure at $y \rightarrow \infty$ can
be represented as
\begin{equation}
p(y) = -\Lambda + \bar{p}(y) \label{infin}
\end{equation}
where $\bar{p}(y)$ is a positive function vanishing at $y
\rightarrow \infty$. Substituting (\ref{infin}) into Eq.
(\ref{TOVy}), one gets an asymptotic equation
\begin{equation}
\frac{d\bar{p}}{dy} = -\frac{\bar{p}}{y}, \label{aslam}
\end{equation}
which solution is
\begin{equation}
\bar{p} = \frac{F}{y} \label{aslam1}
\end{equation}
where $F$ is a positive constant. This constant does not depend on
the value of the parameter $M_2$. Thus, the family of the geometries
with $p \rightarrow -\Lambda$ at $y \rightarrow \infty$ appears to be
described by the three parameters $r_0, F$ and $M_2$. However,
as in the case of the family of geometries described above with
the pressure vanishing at some of $y=y_0$, we are not free in the
choice of the value of $M_2$ after the values $r_0$ and $F$ are
fixed. Indeed, due to the monotonic behavior of the function
$\bar{M}(y)$, at least one value of the parameter $M_2$ corresponds
to a solution $\bar{M}(y)$ satisfying the initial condition
$\bar{M}(1) = r_0/2$. Thus, one has a two-parameter family of
geometries defined by  fixing thee values of $r_0$ and $F$.
These geometries have a singularity of the Schwarzschild type at
$y = \infty$ ($\chi = \pi,~r = 0$) due to the nonvanishing mass
$\bar{M} = -M_2$. They have another curious feature: the
metric coefficient $g_{tt} = e^{\bar{\nu}(\chi)}$ given by the
formula (\ref{nu}) tends to infinity as $p \rightarrow -\Lambda$
and, hence,  intervals of the proper time $d\tau = \sqrt{g_{tt}}
dt$ tend to infinity. So, one has an infinite blue shift effect in
contrast to the well-known red shift effects in the vicinity of
the Schwarzschild and de Sitter horizons.

Summarizing, solutions of the Tolman-Oppenheimer-Volkoff equations
in the presence of the Chaplygin gas have the following curious
features:
\begin{enumerate}
\item
All the spatial sections of the spacetime manifolds (excluding a
special case of the de Sitter spacetime) are closed.
\item
Some geometries have a divergent scalar curvature invariant at
a finite value of $r$.
\item Some geometries manifest an infinite blue-shift effect.
\end{enumerate}

Unfortunately, the relations between the boundary conditions
$p_{b}, M_{b}$, the parameters, characterizing the crossing of the
equator $r_0, B$ and the ``final parameters'' characterizing the
three family of geometries with qualitatively different behaviors
at $\chi > \pi/2$ cannot be found analytically and should be
studied numerically.

\section{The phantom case: $|p| > \rho$}

Now consider the system of equations (\ref{TOV1}), (\ref{mass11})
with the boundary condition
\begin{equation}
p(r_b) < -\Lambda.
\label{phant}
\end{equation}
In this case  $|p| > \rho$ that corresponds to  phantom dark energy
and, in principle, to a possibility of creation of wormholes.
If the condition (\ref{phant}) is satisfied, two cases are possible.

Case A:
\begin{equation}
M(r_b) + 4\pi r_b^3 p(r_b) < 0.
\label{phant1}
\end{equation}
In this case, the pressure $p$ is decreasing and its absolute value
is growing. Correspondingly, the energy density is also decreasing and
hence the mass $M$ grows slower than $r^3$. Then the left-hand side of
the expression (\ref{phant1}) is decreasing, too, so the expressions
$(M + 4\pi r^3 p)$, $(r-2M)$ and $(p^2 - \Lambda^2)$ cannot change their
signs.

We examine three possible subcases:
\\
1) $p$ tends to some finite value $-\infty < p_1 < -\Lambda$ when
$r \rightarrow \infty$;\\
2) $p$ tends to $-\infty$ when $r \rightarrow \infty$;\\
3) $p$ grows indefinitely when $r$ tends to some finite value $r_1$.\\

The subcase 1 cannot take place because  the left-hand side of Eq. (\ref{TOV1})
is regular while its right-hand side diverges as $r^2$ when
$r \rightarrow \infty$.

Likewise, the subcase 2 cannot be realized as well. Indeed, suppose that
$p = -p_1 r^{\alpha}, \alpha > 0, p_1 > 0$ when $r \rightarrow
\infty$.  Then Eq. (\ref{phant1}) becomes
\begin{equation}
 \frac{dp}{dr} = -4\pi p^2 r,
\label{phant2}
\end{equation}
which implies
\begin{equation}
 \alpha -1 = 2\alpha +1
\label{phant3}
\end{equation}
or $\alpha = -2$ which contradicts to the positivity of $\alpha$.

We are left with the subcase 3 which can be realized with
\begin{equation}
p = -\frac{p_1}{r_1 -r},\ p_1 > 0, \ \  {\rm when} \ r \rightarrow r_1.
\label{case3}
\end{equation}
Substituting expression (\ref{case3}) into Eq. (\ref{TOV1}), we have the
following relation:
\begin{equation}
\frac{p_1}{(r_1-R)^2} = \frac{p_1^2}{(r_1-r)^2}\cdot \frac{4\pi r_1^2}{r_1-2M_1},
\label{case31}
\end{equation}
from which we  find the value of parameter $p_1$ as a function of the radius
$r_1$ and the mass $M(r_1) = M_1$:
\begin{equation}
p_1 = \frac{r_1-2M_1}{4\pi r_1^2}.
\label{case32}
\end{equation}
Thus, we obtain a two-parameter family of solutions in which one encounters
a singularity at $r=r_1$ because the scalar curvature $R$ diverges there.

Case B:
\begin{equation}
M(r_b) + 4\pi r_b^3 p(r_b) > 0.
\label{phantB}
\end{equation}
In this case the pressure  grows with $r$. Then, in the right-hand
side of Eq.  (\ref{TOV1}) we have three decreasing positive terms
$(p^2 - \Lambda^2)$, $(M + 4\pi p r^3)$ and $(r-2M)$. The problem is
which of them vanishes before the others, if any.

The above terms cannot simultaneously remain positive as
$r \rightarrow \infty$ because in this case $|p| > \Lambda > \rho$ and the
expression $(M + 4\pi p r^3)$ would unavoidably change sign.

The case when $p = -\Lambda$ while the other two expressions remain
positive is also excluded. Indeed, if $p \rightarrow -\Lambda $ as $r$
approaches some finite value, the left-hand side of Eq. (\ref{pL}) has a
logarithmic divergence while its right-hand side is regular. On the other
hand, if $ p \rightarrow -\Lambda$ as $r \rightarrow \infty $, the expression
$(M + 4\pi p r^3)$ will change its sign.

If $(p^2 - \Lambda^2)$ and $(r-2M)$ vanish at some $r = r_0$, Eq. (\ref{pL})
takes the form
\begin{equation}
d\ln(p^2-\Lambda^2) = -\frac{dr}{r-r_0}
\label{pL1}
\end{equation}
which implies
\begin{equation}
p^2 - \Lambda^2 \sim \frac{1}{r_0 - r}
\label{pL2}
\end{equation}
contradicting the hypothesis.
Thus, the pressure cannot achieve the value $p = -\Lambda$.

As shown in the preceding section, the denominator $(r-2M)$ at the right-hand
side of Eq. (\ref{TOV1})  can only vanish at some $r=r_0$ simultaneously with
$(M + 4\pi p r^3)$ and the pressure at  $r = r_0$ should be equal
$p = -\frac{1}{8\pi r_0^2}$ (equator). Let us prove that it is impossible to
achieve the equator if $p < -\Lambda$.
Indeed, the formula (\ref{mass11})  shows that in this case
\begin{eqnarray}
&&r-2M = r_0 + (r-r_0) -2M(r_0) - 2M'(r_0)\,(r-r_0) \nonumber \\
&&= (r-r_0)
(1-64\pi^2\Lambda^2 r_0^4)
\label{difference}
\end{eqnarray}
as $r \rightarrow r_0$.
The difference $(r-2M)$ should be positive as $r \rightarrow r_0$ from below,
so the expression $(1-64\pi^2\Lambda^2 r_0^4)$ should be negative. However,
at $|p| > \Lambda$ this expression is positive - a contradiction.

Thus, if (\ref{phantB}) is satisfied, $p$ grows until some maximum value
$p_{max} < -\Lambda$ when the expression $(M + 4\pi p r^3)$ changes  sign
while the terms $(p^2-\Lambda^2)$ and  $(r-2M)$ are positive, and we led
back to the case A.

So, we have proved that only two regimes are possible for a star-like object
immersed into the phantom Chaplygin gas. If initial conditions satisfy
(\ref{phant1}), the pressure is decreasing and diverges at some finite value
of $r$. Then the space-time acquires a scalar curvature singularity there.
On the other hand, if initial conditions
satisfy (\ref{phantB}), then the pressure grows with the  $r$ until
some maximum value $p = p_{max}$ where the expression in the
right-hand side of Eq. (\ref{TOV1}) changes  sign. After that we
come back to case A: the pressure decreases and explodes  according
to   (\ref{case3}) at some finite value of the radial coordinate
$r$. No equator and no horizon are attained in the case of the
phantom Chaplygin gas: the quantity $(r -2M)$ is always positive. As
in the case of the non-phantom Chaplygin gas, the relation between
the initial values of the parameters functions $p(r_b)$ and $M(r_b)$
and the parameters $r_1$ and $M(r_1)$ cannot be found analytically.

We can summarize the results of the above considerations  in the following \\
\\
THEOREM. In a static spherically symmetric universe filled with the
phantom Chaplygin gas, the scalar curvature becomes singular at some
finite value of the radial coordinate and the universe is not
asymptotically flat.

\section{Static spherically symmetric universe filled exclusively with
the phantom Chaplygin gas}

Now we study spherically symmetric static solutions for a universe filled
exclusively with the phantom Chaplygin gas. The theorem above is valid in
this case, too. This situation is of much interest because when the weak
energy condition is violated, $\rho + p < 0$, wormholes may appear (though
not necessarily, see \cite{BS2007} in this connection). Suppose that at
some finite value of the radial variable $r = r_b$, the factor
$r_b - 2M(r_b)$ is positive and $p(r_b) < -\Lambda$. Then, as in the
preceding analysis, one can consider the evolution of the functions $M(r)$
and $p(r)$ in accordance with Eqs. (\ref{TOV1}), (\ref{mass11}) but with a
{\it decreasing} value of the radial variable $r$. Now only two
possibilities may be realized: one can arrive at the value $r = 0$ keeping
always a positive value of the factor $r- 2M$, or one can encounter a
situation when at some finite value of $r = r_0$ this factor vanishes.

Consider first the case when  $(r-2M)$ is positive for all  values $r > 0$.
Here, one can imagine two different regimes as $r$ approaches zero. In the
first one the mass in the vicinity of $r = 0$ is positive and behaves as
$M \sim r^{\alpha}, \alpha > 0$. In the second regime the mass tends to a
negative constant when $r \rightarrow 0$. A detailed analysis shows that
only the first regime is compatible with the TOV equations (\ref{TOV1}),
(\ref{mass11}) in the presence of the phantom Chaplygin gas. Precisely, in
the vicinity of $r = 0$ the pressure and the mass functions have the form:
\begin{equation}
p = p_0 -\frac{8\pi^2(3p_0^2 - \Lambda^2)(p_0^2 - \Lambda^2) r^2}{3p_0^2},
\label{pressure-reg}
\end{equation}
\begin{equation}
M = -\frac{4\pi \Lambda^2 r^3}{3p_0}
\label{mass-reg}
\end{equation}
where $p_0$ is an arbitrary number such that $p_0 < -\Lambda$. Being
regular at the center $r = 0$, this static configuration develops a
singularity at some finite value of the radius $r_1$, where the
pressure becomes equal to minus infinity. Thus, we have a
one-parameter family of  static spherically symmetric solutions
of the Tolman-Oppenheimer-Volkoff equations in a world filled with
the phantom Chaplygin gas. This family is parameterized by the value
of the pressure at the center $r = 0$.

Now suppose that the factor $r > 2M(r)$ vanishes at some value $r = r_0$.
An analysis similar to the one carried out in Sec.3 shows that this is
positive only if also the expression $M + 4\pi p r^3$ in the numerator
of the right-hand-side of Eq. (\ref{TOV1})
vanishes at $r_0$. In Sec. 3  the surface $r = r_0$ was
called equator because it corresponded to the maximal value of the
radial variable $r$. Now it corresponds to the minimal value of $r$,
and it is nothing but a throat. Just like in the case of the equator
considered in Sec. 3, the throat can be achieved only at
 $p = -\frac{1}{8\pi r_0^2}$. In the phantom
case $p < -\Lambda$, hence, there is a restriction on the size of
the throat
\begin{equation}
r_0 < \sqrt{\frac{1}{8\pi \Lambda}}. \label{throat}
\end{equation}

In order to  describe the  crossing of the throat, it is convenient
to introduce the hyperbolic coordinate $\eta$ instead of the radius
$r$,
\begin{equation}
r = r_0 \cosh \eta, \label{eta}
\end{equation}
which plays a role similar to that played by the trigonometrical
angle $\chi$ in the description of the equator.

Now the TOV equations look like
\begin{equation}
\frac{dp}{d\eta} = \frac{(\Lambda^2 - p^2)(M + 4\pi p r_0^3\cosh^3
\eta )\sinh \eta}{p\cosh \eta (r_0\cosh \eta - 2 M)},
\label{throat1}
\end{equation}
\begin{equation}
\frac{dM}{d\eta} =
-\frac{4\pi\Lambda^2r_0^3\cosh^2\eta\sinh\eta}{p}. \label{throat2}
\end{equation}
The solution of  (\ref{throat2}) at small values of $\eta$ is
\begin{equation}
M = r_0 + 16\pi^2\Lambda^2 r_0^5 \eta^2~. \label{throat3}
\end{equation}
Representing the pressure as
\begin{equation}
p(\eta) = -\frac{1}{8\pi r_0^2} + \tilde{p}(\eta), \label{throat4}
\end{equation}
one can rewrite Eq. (\ref{throat1})
 in the neighborhood of the throat as
 \begin{equation}
 \frac{d\tilde{p}}{d\eta} = \frac{\tilde{p}}{\eta} + C_T \eta
 \label{throat5}
 \end{equation}
where the negative constant $C_T$ is equal to
\begin{equation}
C_T = \frac{64\pi^2\Lambda^2 r_0^4 - 3}{16\pi r_0^2}~. \label{CT}
\end{equation}
The solution of Eq. (\ref{throat5}) is
\begin{equation}
\tilde{p} = D \eta + \frac12 C_T \eta^2 \label{sol-thr}
\end{equation}
where $D$ is an arbitrary constant.

Thus, geometries with a throat  constitute a two-parameter
family, characterized by the value of the throat radius $r_0$ and by
the value of the parameter $D$. From Eq. (\ref{sol-thr}) we see that
when the hyperbolic parameter $\eta$ grows, the negative term
$\frac12 C_T \eta^2$ starts dominating, while the pressure decreases
and achieves an infinite negative value at a finite value of the
radius $r$ where we encounter a curvature singularity. The
peculiarity of this configuration with a throat consists in the fact
that these singularities are achieved at both sides of the throat
or, in other words, at one positive $\eta_1 > 0$  and one negative
$\eta_2< 0$ values of the hyperbolic parameter. The values $\eta_1$
and $\eta_2$ correspond to  values $r_1 = r_0\cosh\eta_1$ and $r_2 =
r_0\cosh\eta_2$ of the radial parameter. If $D=0$ one has $r_1 =
r_2$. The solutions with a throat could be seen as wormhole-like,
but in contrast with the Morris-Thorne-Yurtsever wormholes
\cite{worm-stand}, they do not connect two asymptotically flat
regions.

We summarize the preceding considerations concerning solutions of
the Tolman-Oppenheimer-Volkoff equations in the universe filled with
the phantom Chaplygin gas as follows. Suppose we start from an
initial condition $\bar{r}-2M(\bar{r}) > 0$ and $p(\bar{r}) <
-\Lambda$ for some given value $\bar{r}$ of $r$. Then letting the
functions $M(r)$ and $p(r)$ evolve to values $r > \bar{r}$, we
unavoidably arrive to an infinite negative value of the pressure at
some finite value $r_f$ of the radius $r$, thus encountering a
singularity. Instead, for $r < \bar{r}$ two qualitatively different
situations may arise: either we arrive to $r = 0$ in a regular way,
or we may discover a throat at some finite value $r = r_0$, this
being the generic situation. Upon passing the throat, an observer
finds itself in another patch of the world and then, with the radius
increasing, stumbles again upon a curvature singularity at some
finite distance from the throat.

Note that in the traditional view of wormholes, one supposes that
there is a minimal value of the radial parameter characterizing a
throat, and the space-time at both sides of the throat is either
asymptotically flat, or has some other traditional structure (for
example, wormholes could also connect two expanding asymptotically
Friedmann universes). Here we have found a different kind of
wormhole-like solutions: those connecting two space-time patches
which have a scalar curvature singularity at some finite value of
$r$.

We conclude this section with the brief comment on the results of
paper \cite{Kuh} where the problem of the existence of wormhole
solutions supported by the phantom Chaplygin gas was studied.  The
main part of this paper is devoted to the consideration of the so
called ``anisotropic'' Chaplygin gas, i.e. a fluid whose radial
pressure satisfies the Chaplygin gas equation of state, while the
tangential pressure can be arbitrary. Then the system of TOV
equations is under-determined and its solution contains one
arbitrary function. Choosing this function in a convenient way, one
can construct a lot of solutions, satisfying the desired properties.
However, this fluid is not the Chaplygin gas and, moreover, is not a
barotropic fluid.  The case of the isotropic Chaplygin gas is also
considered in Ref.\cite{Kuh}. The author studies numerically the
behavior of metric coefficients in the vicinity of the throat
without considering the problem of continuation of this solution to
larger values of the radial coordinate. However, the general theorem
proved at the end of the section IV of the present paper states that
all spherically-symmetric solutions supported by the phantom
Chaplygin gas (with or without a throat) have a curvature
singularity at some finite value of the radial coordinate and hence
cannot be asymptotically flat.

\section{Conclusion}

In this paper we have studied the Tolman--Oppenheimer--Volkoff
equations for static spherically symmetric objects immersed in the
space filled with the Chaplygin gas. Both cases, phantom and
non-phantom, were considered. In the non-phantom case all solutions
(excluding the de Sitter one) represent a spheroidal geometry, where
the radial coordinate achieves a maximal value (equator). After
crossing the equator, depending on the boundary conditions, three
types of solutions can arise: a closed spheroid having a
Schwarzschild-type singularity with  infinite blue-shift at the
"south pole", a regular spheroid, and a truncated spheroid having a
scalar curvature singularity at a finite value of the radial
coordinate.

For the case of the phantom Chaplygin gas, the equator is absent and
all star-like external solutions have the geometry of a truncated
spheroid having a scalar curvature singularity at some finite value
of the radial coordinate. Besides, we have also considered the
static spherically symmetric configurations existing in a universe
filled exclusively with the phantom Chaplygin gas. Here two cases
are possible: geometries which are regular at the center $r = 0$ and
having a scalar curvature singularity at some finite value of $r$,
and  geometries containing throats connecting  two patches of the
world which again have scalar singularities at some finite values of
the radius. Because of these singularities, the construction of
stable, traversable and asymptotically flat wormholes using the
phantom Chaplygin gas is prohibited, in spite of breaking the weak
energy condition in this case.

Finally, we note that many of the solutions of the TOV equations
studied above possess  singularities arising at some finite values
of the radial coordinate, while intuitively it is more habitual to
think about singularities arising at the point characterized by
the vanishing of this coordinate. Something similar happens in the
study of isotropic and homogeneous cosmological models, too.
Here, in addition to the traditional Big Bang and Big Crunch
singularities, an intensive study of the singularities which take
place at finite or at infinite values of the cosmological scale
factor is under way (see e.g. \cite{cosmol}).

\section*{Acknowledgement}
This work was partially supported by the Russian Foundation for Basic research via grant RFBR 08-02-00923,
by the Program of support of Leading Scientific Schools of the President of the Russian Federation via grant 
 LSS-4899.2008.2, and by the Scientific Programme "Astronomy" of
the Russian Academy of Sciences.

\end{document}